\documentclass[10pt, conf, twocolumn]{IEEEtran}
\usepackage{color}
\usepackage{times}
\usepackage{epsfig}
\usepackage{graphicx}
\usepackage{epstopdf}
\usepackage{algorithm}
\usepackage{algorithmic}
\usepackage{amsmath}
\usepackage{amssymb}
\usepackage{amsxtra}
\usepackage{multirow}
\usepackage{mathtools}
\usepackage{caption}
\usepackage{url}
\usepackage{subfigure}
\usepackage{cite}
\usepackage{subfigure}
\usepackage{amsthm}
\newtheorem{theorem}{Theorem}
\newtheorem{lemma}{Lemma}

\newtheorem{assumption}{Assumption}
\newtheorem{remark}{Remark}
\newtheorem*{proof*}{Proof}

\begin{document}

\title{Secure and Reconfigurable Network Design for Critical Information Dissemination in the Internet of Battlefield Things (IoBT)}

\author{ \IEEEauthorblockN{\large Muhammad Junaid Farooq and Quanyan Zhu} \\ \IEEEauthorblockA{Department of Electrical \& Computer Engineering, Tandon School of Engineering, \\New York University, Brooklyn, NY, USA,} Emails: \{mjf514, qz494\}@nyu.edu. \vspace{-0.2in}
}

\maketitle

\begin{abstract}
The Internet of things (IoT) is revolutionizing the management and control of automated systems leading to a paradigm shift in areas such as smart homes, smart cities, health care, transportation, etc. The IoT technology is also envisioned to play an important role in improving the effectiveness of military operations in battlefields. The interconnection of combat equipment and other battlefield resources for coordinated automated decisions is referred to as the Internet of battlefield things (IoBT). IoBT networks are significantly different from traditional IoT networks due to the battlefield specific challenges such as the absence of communication infrastructure, and the susceptibility of devices to cyber and physical attacks. The combat efficiency and coordinated decision-making in war scenarios depends highly on real-time data collection, which in turn relies on the connectivity of the network and the information dissemination in the presence of adversaries. This work aims to build the theoretical foundations of designing secure and reconfigurable IoBT networks. Leveraging the theories of stochastic geometry and mathematical epidemiology, we develop an integrated framework to study the communication of mission-critical data among different types of network devices and consequently design the network in a cost effective manner.
\end{abstract}

\IEEEpeerreviewmaketitle


\vspace{-0.1in}
\section{Introduction}
\begin{figure}
\centering
\includegraphics[width=3.1in]{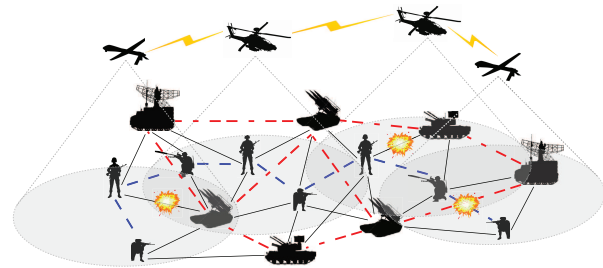}\\
\caption{A typical IoBT network with heterogeneous battlefield things and random enemy attacks. Battlefield devices interact with other devices within their communication range in a D2D manner for exchange of mission critical information. The link shape illustrates the piece of information being shared.
}\label{iobt_diagram}
\end{figure}

The Internet of things (IoT) is an emerging paradigm that allows the interconnection of devices which are equipped with electronic sensors and actuators~\cite{iot_ref1}. It allows for a higher level of situational awareness and effective automated decisions without human intervention. The concept has proven to be extremely useful in applications such as smart homes, energy management, smart cities, transportation, health care and other areas~\cite{iot_ref2}. Recently, there is an interest in the defence community to leverage the benefits enabled by the IoT to improve the combat efficiency in battlefields and effectively manage war resources. This emerging area of using IoT technology for defence applications is being referred to as the Internet of battlefield things (IoBT)~\cite{iobt_ref2}. Fig.~\ref{iobt_diagram} illustrates a typical battlefield comprising of heterogeneous objects, such as soldiers, armoured vehicles, and aircrafts, that communicate with each other amidst cyber-physical attacks from the enemy.

The IoBT has the potential to completely revolutionize modern warfare by using data to improve combat efficiency as well as reduce damages and losses by automated actions while reducing the burden on human war-fighters. Currently, the command, control, communications, computers, intelligence, surveillance and reconnaissance (C$^4$ISR) systems use millions of sensors deployed on a range of platforms to provide situational awareness to military commanders and troops, on the ground, seas, and in the air~\cite{command_control}. However, the real power lies in the interconnection of devices and sharing of sensory information that will enable humans to make useful sense of the massive, complex, confusing, and potentially deceptive ocean of information. In the battlefield scenarios, the communications between strategic war assets such as aircrafts, warships, armoured vehicles, ground stations, and soldiers can lead to improved coordination, which can be enabled by the IoBT~\cite{iobt_ref1}. However, to become a reality, this vision will have to overcome several technical limitations of current information systems and networks.

Most civilian IoT applications such as smart homes and cities are infrastructure based, where the devices are connected to each other and the internet via an access point or gateway. In the battlefield scenario, the communication infrastructure such as cellular networks or base stations may not be available. Hence, the \emph{battlefield things} need to exploit device-to-device (D2D) communications~\cite{d2d_iot} to share information with other things\footnote{We use the terms ``things" and ``devices" interchangeably to refer to battlefield things throughout the paper.}. Therefore, the information sharing can be affected by the physical parameters of the network such as the transmission power of the things, the number of deployed things, their locations, and the flexibility of communication between different types of things. In addition to these factors, another impediment in the connectivity of battlefield things is the susceptibility to cyber-physical attacks. The information exchange between things may be affected by several factors such as jamming of radio frequency channels, physical attacks on infrastructure, node failures due to attacks on power sources, or lack of power, etc. Since the analytics and automated decisions in an IoBT network rely on the real-time data provided by the sensors deployed in the battlefield, we need to ensure the spread of information in the networks with a certain level of reliability and security to make accurate decisions.

Although the IoBT has to do to with much more beyond the connectivity of battlefield things, such as digital analytics and automated response, which allows the systems to react more quickly and precisely than humans; however, the connectivity aspect is vital in allowing the IoBT systems to unleash their full potential. It is ideal if we can achieve complete situational awareness and perfect information spreading throughout the network. However, due to limited available resources, incurred costs (capital and operational), and vulnerability to attacks, it is not practical and hence requires a cost-effective, secure and reconfigurable network design. Therefore, in this paper, we first characterize the information dissemination in an IoBT network and use it to design the network parameters to achieve mission specific performance goals with minimal amount of resources.

Stochastic geometry (SG) based models have been successfully used in the modeling and analysis of traditional wireless networks such as cellular networks~\cite{andrews_coverage} and ad-hoc wireless sensor networks~\cite{haenggi_sg}. These models accurately capture the effect of spatial distribution of network devices and characterize the resulting performance.
On the other hand, epidemic models~\cite{epidemiology} have been studied extensively for analyzing the spread of viruses in computers, rumours in humans, and infectious diseases in biological networks~\cite{viruses,wireless_epidemics1,wireless_epidemics2}. Although these models are highly useful in capturing certain aspects of the networks, none of them captures the unique characteristics of IoBT networks. The SG models lack the capability of analysing the dynamics of information dissemination in networks. While a percolation study for SG models is available to determine the connectivity of the network (as shown in~\cite{haenggi_sg}), however, it does not capture the dynamics of information dissemination and the effects of cyber-physical attacks. On the other hand, epidemic models fail to incorporate the geometry of the network and hence cannot give meaningful insights in physically deployed communication networks. Moreover, there are few descriptive models available in literature for designing IoT networks, most of which are developed for civilian applications~\cite{hesham_iot} and do not incorporate the ad hoc nature of IoT networks over battlefields. Hence, it is imperative to develop an integrated design framework that can capture the unique characteristics of IoBT networks.

In this paper, we develop a SG based model to characterize the connectivity of IoBT networks in terms of the degree distribution. We then use an epidemic spreading model to capture the dynamic information dissemination within and across networks of devices at the equilibrium state. The resulting integrated open-loop system model is used as a basis for reconfiguring the network parameters to ensure a mission-driven information spreading profile in the network.

\vspace{-0.0in}
\section{System Model}
In this section, we first describe the geometry of the IoBT network and propose an abstraction model using tools from stochastic geometry. Then, we model the spread of information in the heterogeneous IoBT network inspired from mathematical epidemiology.

\vspace{-0.1in}
\subsection{Network Geometry}
\begin{figure}
  \centering
  \includegraphics[width=3.2in]{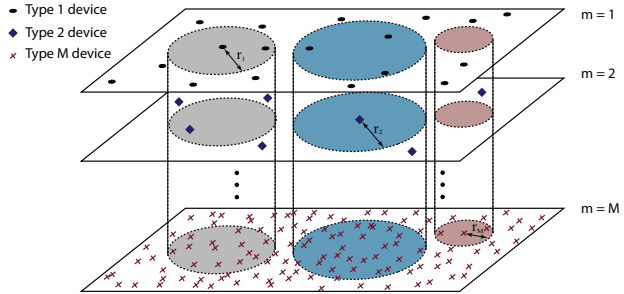}\\
  \caption{Heterogeneous IoBT network decomposed into virtual layers, each containing devices of the same type.}\label{sys_model}
\end{figure}

We consider a heterogeneous adhoc wireless network composed of $M$ different types of devices. Each device corresponds to a different battlefield thing such as a soldier equipped with smart devices, armoured vehicle, ground station, unmanned aerial vehicle (UAV), etc. The different types of devices are characterized by their transmission power or equivalently, the communication range $r_m \text{ in meters}$ and the uniform deployment density in $\mathbb{R}^2$ denoted by $\lambda_m$ devices per km$^2$, $\forall m = 1,\ldots,M$. The communication range of the devices is tunable in the interval $[r_m^{\min}$,$r_m^{\max}]$, where $r_m^{\min} \geq 0$ and $r_m^{\max} \geq r_m^{\min}, \forall m = 1, \ldots,M$. The devices of type $m$ can be abstracted as a homogeneous Poisson Point Process\footnote{The PPP assumption reflects the lack of structure in the spatial distribution of the nodes and is appropriate to use in the case of a large number of nodes where it is difficult to keep track of the topology.} (PPP)~\cite{sg} of intensity $\lambda_m$, denoted by $\Phi_m$. Assuming that each type of device is placed independently of the other types, the combined IoBT network can be represented as a PPP of intensity $\Lambda$\footnote{This result follows from the superposition theorem of PPPs~\cite{sg}.}, denoted by $\Phi$, where $\Lambda = \sum_{m=1}^{M}\lambda_m$. Due to the absence of traditional communication infrastructure such as base stations, the devices are only able to communicate using D2D communications, i.e., device $x_m$ of type $m$ can communicate with a device $y_n$ of type $n$ only if $\|x_m - y_n \| \leq r_m$ and vice versa, where $\|.\|$ represents the Euclidean distance. Hence, the communication links between devices can be modeled using a random geometric graph (RGG)~\cite{rgg} with a given connection radius. For the ease of exposition of the network and connectivity between different type of devices, we virtually decompose the network into $M$ layers, where each layer contains a different type of device. An illustrative representation of the network model is provided in Fig.~\ref{sys_model}. The connectivity between things of the same type, labeled as \emph{intra-layer} connectivity, the connectivity between different type of things, labeled as \emph{inter-layer} connectivity, and the \emph{combined network} connectivity are explained in the subsequent subsections.

\subsection{Network Connectivity}\label{connectivity_sec}
In this subsection, we describe the connectivity between the heterogeneous things in an IoBT network. The connectivity of devices can be classified into \emph{intra-layer}, \emph{inter-layer}, and \emph{combined} network connectivity, which are explained as follows:
\subsubsection{Intra-layer Connectivity}
Within a particular network layer $m$, devices can communicate with each other if they are within a distance of $r_m$ of each other. The set of communication neighbours of a typical device of type $m$, referred to as $x_m$, can be expressed as $\mathcal{N}_m(x_m) = \{y_m \in \Phi_m : \|x_m-y_m\| \leq r_m\}$. The connectivity of the RGG formed by devices of type $m$ is characterized by the degree of the devices, denoted by $K_m$, which is defined as the average number of neighbours of each device, i.e., $K_m = |\mathcal{N}_m(x_m)|$, where $|.|$ represents the set cardinality. Due to the PPP assumption, the intra-layer degree, is a Poisson random variable in the mean field, and can be expressed by the following lemma:
\begin{lemma} \label{lemma_intra_layer_degree}
The intra-layer degree distribution of network layer $m$ can be expressed as follows:
\begin{align}
\mathbb{P}(K_m = k) = e^{-\lambda_m \pi r_m^2}\frac{(\lambda_m \pi r_m^2)^k}{k!}, \ k \geq 0,
\end{align}
for sufficiently large $\lambda_m$ and $r_m$. The average intra-layer degree of layer $m$ can be expressed as $\mathbb{E}[K_m] = \lambda_m \pi r_m^2$.
\end{lemma}
From Fig.~\ref{sys_model}, it is easy to see that the average degree, or equivalently the connectivity, depends on the density of the deployed devices as well as the communication range.

\subsubsection{Inter-layer Connectivity}
The devices in one network layer can communicate with the devices in other network layers that are inside their region of influence. The region of influence of a typical node at each layer is illustrated by the projected circles as shown in Fig.~\ref{sys_model}. The set of devices of layer $n$ under the influence of a typical device at layer $m$ can be expressed as $\mathcal{N}_{mn}(x_m) = \{y \in  \{\Phi_m \cup \Phi_n\} : \|x_m-y\| \leq r_m\}$. The inter-layer degree between layer $m$ and $n$, denoted by $K_{mn}$\footnote{With a slight abuse of notation, we denote $K_m$ as the average intra-layer degree of devices in layer $m$ and $K_{mn}$ as the average inter-layer degree between layer $m$ and $n$.}, can be expressed as $K_{mn} = |\mathcal{N}_{mn}(x_m)|$. Note that the inter-layer degree is not symmetric, i.e., $K_{mn} \neq K_{nm}$. We assume reciprocity of communication between different devices which is formally stated in Assumption 1.
\begin{assumption}
We assume that the devices that are under the influence of another type of device can also communicate back with that device. In practice, this can be realized using different radio interfaces or protocols for communicating between different types of devices.
\end{assumption}
The inter-layer degree distribution is expressed by the following lemma.
\begin{lemma}\label{interlayer_degree_lemma}
The probability distribution of the inter-layer degree between devices of layer $m$ and $n$, denoted by $K_{mn}$, can be written as follows:
\begin{align}
\mathbb{P}(K_{mn} = k) =  e^{- (\lambda_m + \lambda_n)\pi r_m^2} \frac{((\lambda_m + \lambda_n) \pi r_m^2)^k}{k!}, \ k \geq 0,
\end{align}
and the average inter-layer degree can be expressed as $\mathbb{E}[K_{mn}] = (\lambda_m + \lambda_n) \pi r_m^2$.
\end{lemma}

\subsubsection{Combined Network Connectivity}
The total network connectivity is characterized by the average degree of the combined network. In the combined network, where all types of devices are able to communicate with each other using Assumption 1, the degree of each device is evaluated to be the total number of devices of all types inside its area of influence. The average total network degree is distributed as a multi-modal Poisson random variable and can be expressed by the following lemma.
\begin{lemma}\label{combined_degree_lemma}
The degree of the combined network, denoted by $K_{o}$, is distributed as follows:
\begin{align}
\mathbb{P}(K_{o} = k) = \frac{1}{\Lambda} \sum_{m=1}^{M} \lambda_m e^{- \Lambda \pi r_m^2} \frac{(\Lambda \pi r_m^2)^k}{k!}, \ k \geq 0,
\end{align}
where $\Lambda = \sum_{m=1}^M \lambda_m$. The average degree of the combined network can be expressed as $\mathbb{E}[K_o] = \sum_{m=1}^{M} \lambda_m \pi r_m^2$.
\begin{proof*}
See \textbf{Appendix~\ref{combined_degree_lemma_proof}}
\end{proof*}
\end{lemma}

\vspace{-0.0in}
\subsection{Information Dissemination}
Each type of device in the IoBT network generates data which need to be propagated to other devices of the same type and/or different types of devices depending on the role of that device. There are certain pieces of information that needs to be shared among the same type of devices, e.g., soldiers need to communicate certain information with other soldiers. On the other hand, there are pieces of information that needs to be propagated from one network to the other, e.g., sensors transmitting data to a control network. Moreover, some information might be important for all network nodes such as network health monitoring data or network discovery beacons. We assume a time slotted system, in which the devices communicate with other devices for a duration of $\tau$ s. The effective information spreading rate between any two devices within the communication range, denoted by $\alpha$, can be expressed as:
\begin{align}
\alpha = \gamma \times (1 - \delta),
\end{align}
where $\gamma$ is the contact rate, i.e., the total number of transmission attempts made in the interval $\tau$, and $\delta$ is the probability of unsuccessful transmission due to cyber-physical attacks, which we will refer to as \emph{threat level}. Without loss of generality, we can select the contact rate $\gamma = 1$, so effectively, $\alpha = 1 - \delta$ is the probability of successful information transmission between devices\footnote{There is no loss of generality since $\tau$ can be made arbitrarily small.}. We assume a uniform threat level for all network devices. Characterizing the threat level in battlefield scenarios due to jamming and physical attacks, or other adversarial actions has been explored in literature, e.g.,~\cite{jamming} and~\cite{attacks2}, and is beyond the scope of the current work. The dynamics of information dissemination under a given threat level can be described as follows:

\subsubsection{Information Dynamics}
The devices in the network spread information from one device to another in a broadcast manner in each time slot. This process is repeated in all the time slots and the devices can either be in an uninformed state or an informed state depending on the success of information delivery. To model this behaviour and explain the dynamics of information dissemination across the IoBT network, we use the susceptible-infected-susceptible (SIS) model~\cite{epidemics}, which is well studied in mathematical epidemiology. The information dissemination in the network is directly related to the average degree of the network, as described in Section~\ref{connectivity_sec}, which in turn depends on the physical network parameters. Since the network is random with potentially large number of devices, we use the degree based mean-field approach, in which all devices are considered to be statistically equivalent in terms of the degree and the analysis is done on a typical device. The information dissemination dynamics for a typical device can therefore be written as follows~\cite{contact_process}:
\begin{align} \label{differential_equation}
\frac{d I^{(i)}_{k}(t)}{dt} = - I^{(i)}_{k}(t) + \alpha k (1 - I^{(i)}_{k}(t)) \Theta^{(i)}(t),
\end{align}
where $I^{(i)}_k(t)$ denotes the density of informed devices at time $t$ with degree $k$ and information strand $i \in \{m, mn, o\}, \ \forall m, n = 1, \ldots,M$.
\begin{remark}
A strand $m$ refers to a message or piece of information propagating in the network of devices of type $m$. By extension, strand $mn$ refers to the inter-layer information between devices of type $m$ and $n$ that originates from layer $n$. Finally, strand $o$ refers to the global information that is shared by all devices.
\end{remark}

The first term in \eqref{differential_equation} explains the annihilation of information with time, i.e., the informed devices return to the uninformed state at a rate of unity. The second term accounts for the creation of informed nodes due to the spreading. The rate of increase in the density of informed nodes with degree $k$ is directly proportional to the degree, the probability of successful transmission of information $\alpha$, the probability that the node with degree $k$ is not informed, i.e., $(1 - I_{k}^{(i)}(t))$, and the average probability that a neighbour of a device with degree $k$ is informed, denoted by $\Theta^{(i)}(t)$. In our case, since the network is PPP, i.e., uncorrelated, $\Theta^{(i)}(t)$ can be expressed as follows:
\begin{align} \label{theta_expression}
\Theta^{(i)}(t) = \sum_{k \geq 0} \frac{k \mathbb{P}(K_i = k)}{\mathbb{E}[K_i]} I^{(i)}_{k}(t), \ i \in \{m, mn, o\},
\end{align}
where $\mathbb{P}[K_i = k]$ and $\mathbb{E}[K_i]$ are evaluated in Section~\ref{connectivity_sec}.
\subsubsection{Steady State Analysis}
We are interested in determining the steady state of the information dissemination. To this end, we impose the stationarity condition, i.e., set $\frac{d I^{(i)}_{k}(t)}{dt} = 0$. It results in the following expression:
\begin{align} \label{I_k_expression}
I^{(i)}_k = \frac{\alpha k \Theta^{(i)}(\alpha)}{1 + \alpha k \Theta^{(i)}(\alpha)}, \ i \in \{m, mn, o\}.
\end{align}
Notice that $\Theta^{(i)}(\alpha)$ is now a constant that depends on $\alpha$. Now, \eqref{theta_expression} and \eqref{I_k_expression} present a system of equations that needs to be solved self-consistently to obtain the solution for $\Theta^{(i)}(\alpha)$ and $I^{(i)}_k$. In the subsequent section, we deal with the solution of the dynamical information spreading process for the IoBT network.

\vspace{-0.0in}
\section{Methodology}
In this section, we first present a solution to the dynamical information spreading system in IoBT networks and then use it for the efficient design of IoBT networks for mission-specific battlefield applications.

\vspace{-0.0in}
\subsection{Equilibrium Analysis}
Equilibrium analysis provides us with the steady-state situation of the information in the network. Although, with the changes in network topology and other network configurations, the actual information spread might be different; however, the equilibrium state provides us with a reasonable understanding of the system behavior. In order to find the equilibrium solution, we need to solve the self-consistent system expressed in \eqref{theta_expression} and \eqref{I_k_expression}. In fact, it reduces to obtaining a solution to the following fixed-point system:
\begin{align} \label{fixed-point-equation}
\Theta^{(i)}(\alpha) = \frac{1}{\mathbb{E}[K_i]} \sum_{k\geq 0} k \mathbb{P}(K_i = k) \frac{\alpha k \Theta^{(i)}(\alpha)}{1 + \alpha k \Theta^{(i)}(\alpha)},
\end{align}
for $i \in \{m,mn,o\}, \ \forall m,n = 1,\ldots,M$. An obvious solution for this fixed-point system is $\Theta^{(i)}(\alpha) = 0$; however, it is trivial. It can be shown that \eqref{fixed-point-equation} may have at least one solution in the domain $\Theta^{(i)}(\alpha) > 0$ depending on the value of $\alpha$ (See \textbf{Appendix~\ref{uniqueness}}). The condition for this bifurcation to hold is $\alpha \geq \frac{\mathbb{E}[K_i]}{\mathbb{E}[K_i^2]}$. We show that this bifurcation point is unique in the domain $0 < \Theta^{(i)}(\alpha) \leq 1$ (See \textbf{Appendix~\ref{existence}}). Obtaining this solution in closed form for a PPP setting is not always possible due to the complicated form of $\mathbb{P}(K_i = k)$. Hence, an approximate solution can be obtained using the following theorem:
\begin{theorem} \label{theorem_solution}
If a non-zero solution exists for the information spreading dynamics in \eqref{theta_expression} and \eqref{I_k_expression}, i.e., $\alpha \geq \frac{\mathbb{E}[K_i]}{\mathbb{E}[K_i^2]}$, then for $\mathbb{E}[K_i] \gg 1$, a lower bound approximation of the average probability that a neighbour of a device with degree $k$ is informed, can be expressed as follows:
\begin{align}
\hat{\Theta}^{(i)}(\alpha) \approx \max \left(0, 1 - \frac{1}{\alpha \mathbb{E}[K_i]}\right), \label{theta_approx}
\end{align}
\begin{proof*}
See \textbf{Appendix~\ref{solution}}.
\end{proof*}
\end{theorem}
As shown in Appendix~\ref{solution}, the solution is a lower bound for the true solution and becomes a tight approximation for $\mathbb{E}[K_i]~\gg~1$. In the IoBT network, the physical interpretation of $\mathbb{E}[K_i]$ is the average number of communication neighbours of a device related to information strand $i \in \{m,mn,o\}$. It is reasonable to assume that $\mathbb{E}[K_i] \gg 1$ due to the potential high density of devices in IoBT networks and hence, the solution presented in Theorem~\ref{theorem_solution} is indeed a good approximation to the true solution. The corresponding solution for the density of informed devices $\hat{I}^{(i)}_k$ can be obtained by substituting \eqref{theta_approx} into \eqref{I_k_expression}.
The average density of informed devices with information strand $i$ can then be evaluated as:
\begin{align} \label{avg_density}
\hat{I}^{(i)} = \sum_{k \geq 0} \hat{I}_k^{(i)} \mathbb{P}(K_i = k).
\end{align}

\vspace{-0.0in}
\subsection{Secure and Reconfigurable Network Design}
Once the equilibrium point for the information dissemination has been determined, the next step is to design the IoBT network to achieve mission specific goals while efficiently using battlefield resources. In essence, the network design implies tuning the knobs of the network, which in the case of IoBT networks are the transmission ranges and the node deployment densities of the different types of battlefield things. The problem is eventually to find the modes of the intra-layer and iter-layer degree distributions of the network. The objective is to ensure a certain information spreading profile in the network while deploying the minimum number of devices and using the minimum transmit power. Let $\boldsymbol{\lambda} = [\lambda_1 \ \lambda_2 \ \ldots \ \lambda_M]^T$ represent the vector of device deployment densities and $\mathbf{r} = [r_1 \ r_2 \ \ldots \ r_M]^T$ be the vector of communication ranges of each of the devices in the IoBT network. The minimum density of each device in the network, determined by the mission requirements, is denoted by $\boldsymbol{\lambda}^{\min} = [\lambda_1^{\min} \ \lambda_2^{\min} \ \ldots \ \lambda_M^{\min}]^T$, $\lambda_m^{\min} \geq 0, \ \forall m = 1, \ldots, M$. The maximum deployment density of each device, defined by the capacity of the available devices, is denoted by $\boldsymbol{\lambda}^{\max} = [\lambda_1^{\max} \ \lambda_2^{\max} \ \ldots \ \lambda_M^{\max}]^T$, $\lambda_m^{\max} \geq \lambda_m^{\min}, \ \forall m =1,\ldots, M$. Similarly, the tunable transmission range limits of the devices can be expressed as $\mathbf{r}^{\min} = [r_1^{\min} \ r_2^{\min} \ \ldots \ r_M^{\min}]^T$, $r_m^{\min} \geq 0, \ \forall m = 1,\ldots,M$, and $\mathbf{r}^{\max} = [r_1^{\max} \ r_2^{\max} \ \ldots \ r_M^{\max}]^T$, $r_m^{\max} \geq r_m^{\min}, \ \forall m = 1, \ldots, M$. If $\mathbf{w} = [w_1 \ w_2 \ \ldots \ w_M]^T$ such that $\sum_{m=1}^{M} w_m = 1$ represents the weight vector corresponding to the relative capital cost of deploying each type of device, and $p$ represents the unit operational power cost signifying the importance of network power consumption, then a cost function for the network with densities $\boldsymbol{\lambda}$ and transmission ranges $\mathbf{r}$ can be expressed as follows:
\begin{align}
c(\boldsymbol{\lambda},\mathbf{r}) = \sum_{m=1}^{M} w_m \lambda_m \mathcal{A} + p \sum_{m=1}^{M} \lambda_m \mathcal{A} r_m^\eta,
\end{align}
where $\mathcal{A}$ represents the area of the battlefield in km$^2$ and $\eta$ denotes the path-loss exponent\footnote{The power consumption of a device of type $m$ is proportional to $r_m^{\eta}$.}. The first term represents the total deployment cost of all the network devices while the second term represents the total energy cost of operating all the devices with transmission range $\mathbf{r}$. The weights $\mathbf{w}$ can depend on several factors such as the time required for deployment, the monetary cost involved, or the number of devices available in stock, etc. We can then formulate the secure and reconfigurable network design problem as follows:
\begin{align}
& \underset{\boldsymbol{\lambda},\mathbf{r}}{\text{minimize}}
& & c(\boldsymbol{\lambda},\mathbf{r}) \label{org_objective}\\
& \text{subject to}
& & I^{(m)} \geq T_m, m = 1,\ldots,M, \label{org_const1}\\
&&& I^{(mn)} \geq T_{mn}, (m,n) = 1,\ldots,M, \label{org_const2}\\
&&& I^{(o)} \geq T_{o}, \label{org_const3}\\
&&& \boldsymbol{\lambda}^{\min} \leq \boldsymbol{\lambda} \leq \boldsymbol{\lambda}^{\max}, \mathbf{r}^{\min} \leq \mathbf{r} \leq \mathbf{r}^{\max},\label{org_const4}
\end{align}
where $T_i \in [0,1]$, $i \in \{m,mn,o\}$, are the desired mission-specific intra-layer, inter-layer, and global network information spreading thresholds. Since the knobs for certain types of devices in the network may not be tunable, we do not have complete freedom in selecting the parameters to minimize the cost function. This constrained action space can be incorporated into the optimization by setting $\lambda_m^{\min} = \lambda_{m}^{\max}$ or $r_m^{\min} = r_{m}^{\max}$ for any type $m$ device for which the parameter is not tunable. Since computing $\hat{I}^{(i)}$ in \eqref{avg_density} in closed form is not possible due to the form of $\mathbb{P}(K_i = k)$ for the considered network, obtaining a solution of the constrained optimization problem in \eqref{org_objective}-\eqref{org_const4} is intractable. Therefore, we propose a sub-optimal approach to avoid this intractability while still yielding a plausible solution. Instead of ensuring that the average densities of informed devices $\hat{I}^{(i)}$ exceeds the respective thresholds $T_i$, $i \in \{m,mn,o\}$, we impose a constraint on the densities of informed devices that possess a degree equal to the average degree of the network. In other words, we ensure that $\hat{I}^{(i)}_{\mathbb{E}[K_i]} \geq T^{\prime}_{i}$ for some $T_i^{\prime} < T_i, i \in \{m, mn, o\}$. It is reasonable because the proportion of devices with the mean degree contribute the most in the average information spreading. The resulting problem, after substituting the required expressions from Lemma~\ref{lemma_intra_layer_degree}, \ref{interlayer_degree_lemma}, and \ref{combined_degree_lemma}, simplifies to the following:
\begin{table}[]
\centering
\caption{Physical Parameter Ranges.}
\label{parameters}
\def\arraystretch{1.5}
\setlength\tabcolsep{5pt}
\begin{tabular}{|c|c|c|c|c|c|c|c|}
\hline
\multicolumn{4}{|c|}{Deployment Density (km$^{-2}$)} & \multicolumn{4}{c|}{Transmission Range (m)} \\ \hline
$\lambda_1^{\min}$      & $\lambda_1^{\max}$     & $\lambda_2^{\min}$     & $\lambda_2^{\max}$     & $r_1^{\min}$    & $r_1^{\max}$    & $r_2^{\min}$    & $r_2^{\max}$   \\ \hline
0.1           & 10          & 1          & 40          &     100      &    1000       &     10      &    500      \\ \hline
\end{tabular}
\vspace{-0.1in}
\end{table}
\begin{align}
& \underset{\boldsymbol{\lambda},\mathbf{r}}{\text{minimize}}
& &  c(\boldsymbol{\lambda},\mathbf{r}) \\
& \text{subject to}
&& \lambda_m \pi r_m^2 \geq \frac{1}{\alpha(1 - T^{\prime}_m)}, \; \forall m = 1, \ldots, M, \\
&&& \hspace{-0.6in}(\lambda_m + \lambda_n) \pi r_m^2 \geq \frac{1}{\alpha(1 - T^{\prime}_{mn})}, \; \forall (m,n) = 1,\ldots,M,\\
&&& \sum_{m=1}^{M} \lambda_m \pi r_m^2 \geq \frac{1}{\alpha (1 - T^{\prime}_o)}, \\
&&& \boldsymbol{\lambda}^{\min} \leq \boldsymbol{\lambda} \leq \boldsymbol{\lambda}^{\max}, \mathbf{r}^{\min} \leq \mathbf{r} \leq \mathbf{r}^{\max}.
\end{align}

Note that the objective and constraints are biconvex in the feasible solution space, i.e., $\boldsymbol{\lambda}^{\min} \leq \boldsymbol{\lambda} \leq \boldsymbol{\lambda}^{\max}, \mathbf{r}^{\min} \leq \mathbf{r} \leq \mathbf{r}^{\max}$ with $\boldsymbol{\lambda}^{\min} \geq 0$ and $\mathbf{r}^{\min} \geq 0$. Hence, the problem can be solved using constrained biconvex programming techniques~\cite{biconvex}. The battlefield area $\mathcal{A}$ is a common factor in the objective function and can be safely removed from the optimization problem.

\begin{figure*}[t!]
\addtolength{\subfigcapskip}{-0.0in}
\begin{center}
\subfigure[Transmission ranges against threat level.]{\label{intel_range}\includegraphics[width=2.2in]{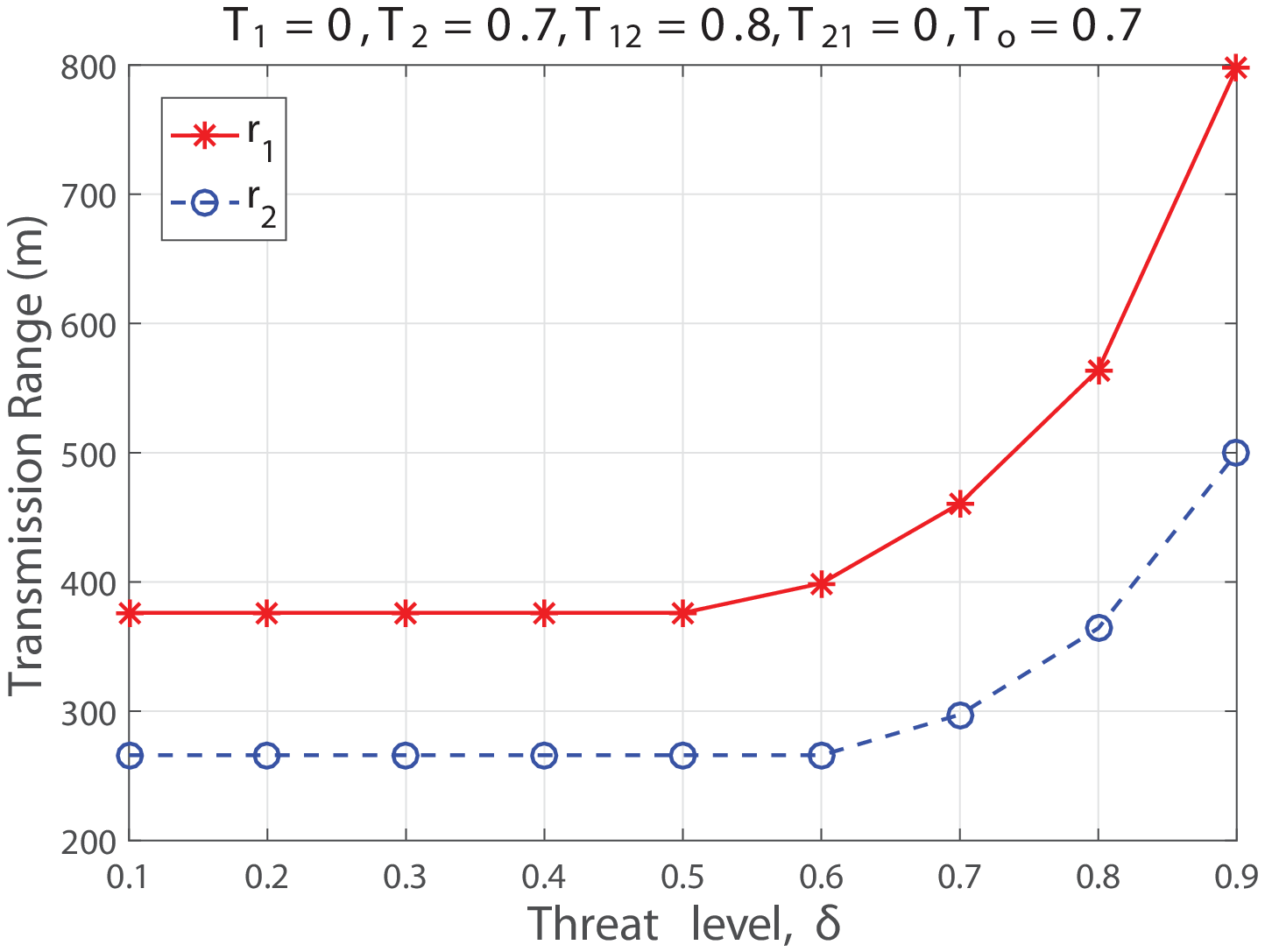}}
\subfigure[Deployment densities against threat level.]{\label{intel_density}\includegraphics[width=2.2in]{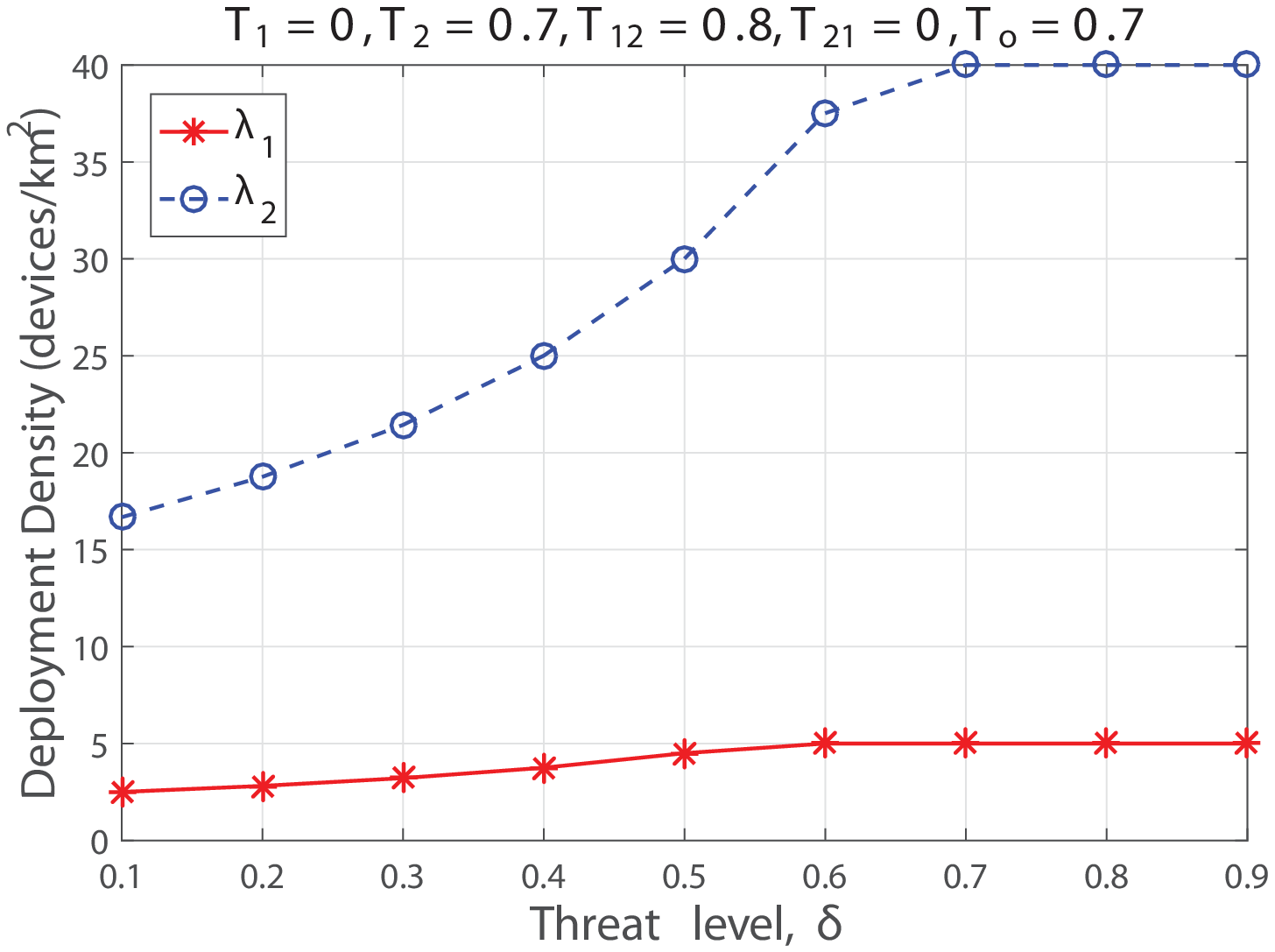}}\
\subfigure[Cost function against threat level.]{\label{intel_cost}\includegraphics[width=2.2in]{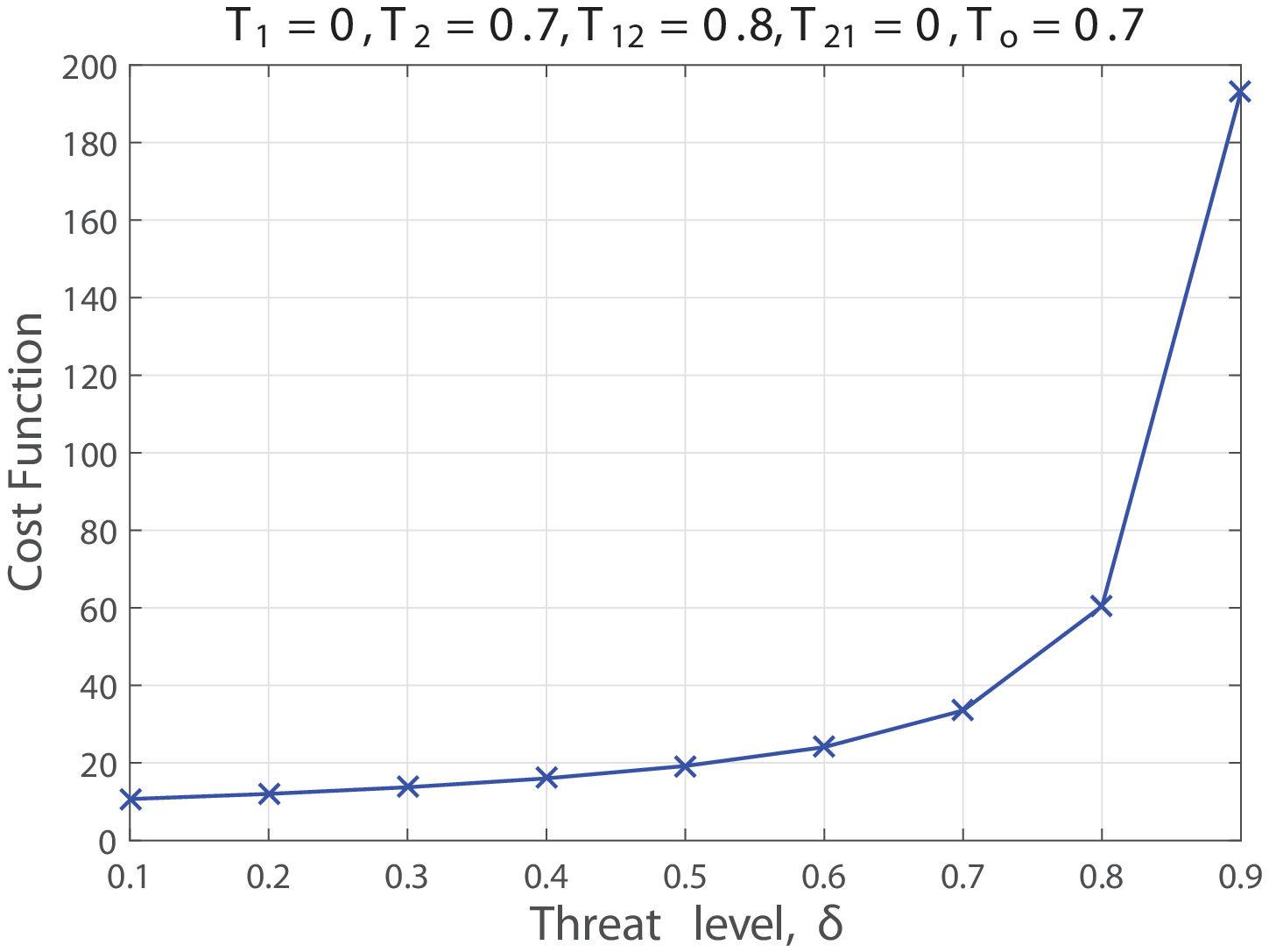}}
\end{center}\vspace{-0.0in}
\caption{Optimal network parameters for the intelligence mission.}
\label{Fig_intelligence}
\end{figure*}
\begin{figure*}[h!]
\addtolength{\subfigcapskip}{-0.0in}
\begin{center}
\subfigure[Transmission ranges against threat level.]{\label{enc_range}\includegraphics[width=2.2in]{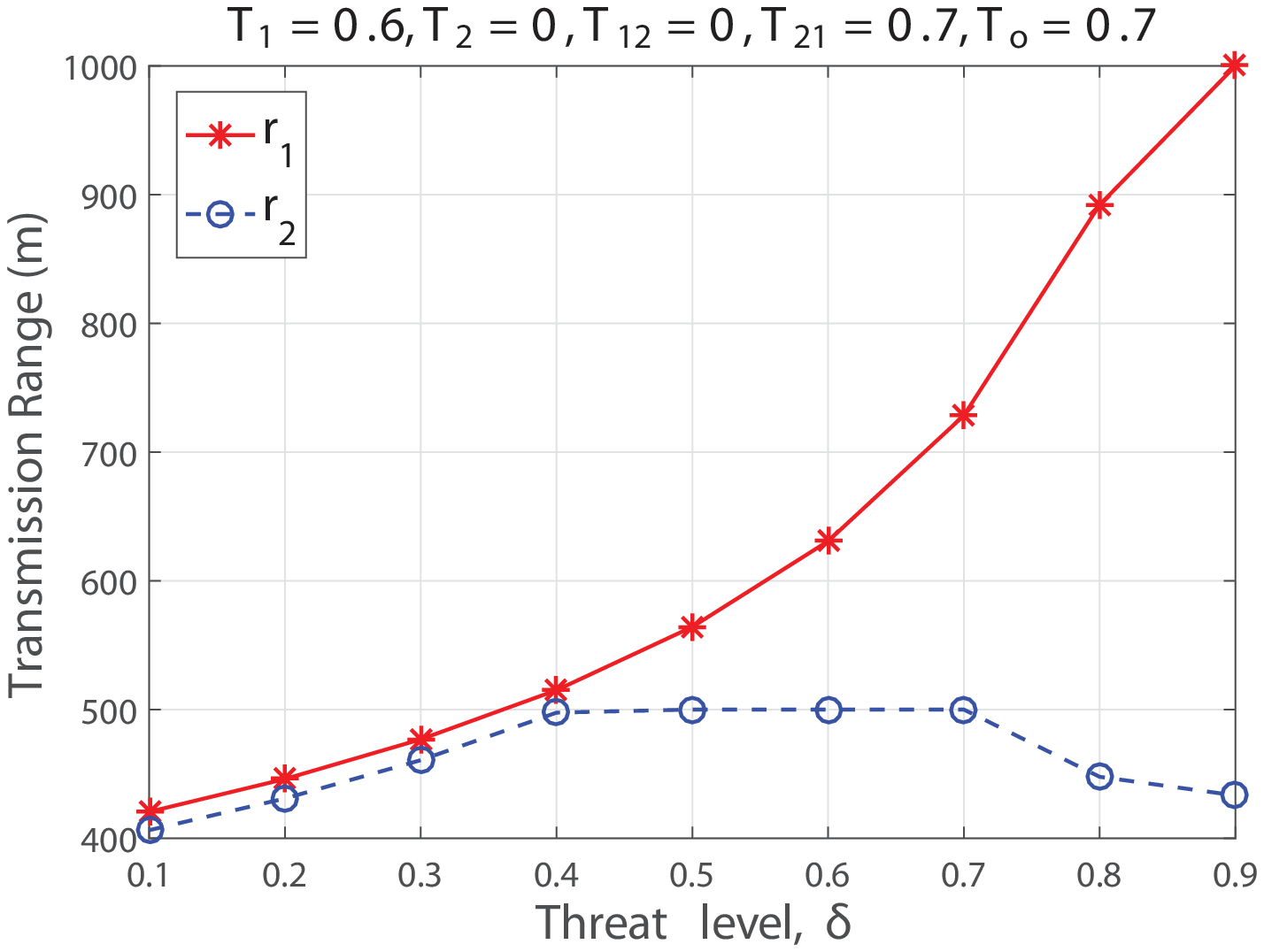}}
\subfigure[Deployment densities against threat level.]{\label{enc_density}\includegraphics[width=2.2in]{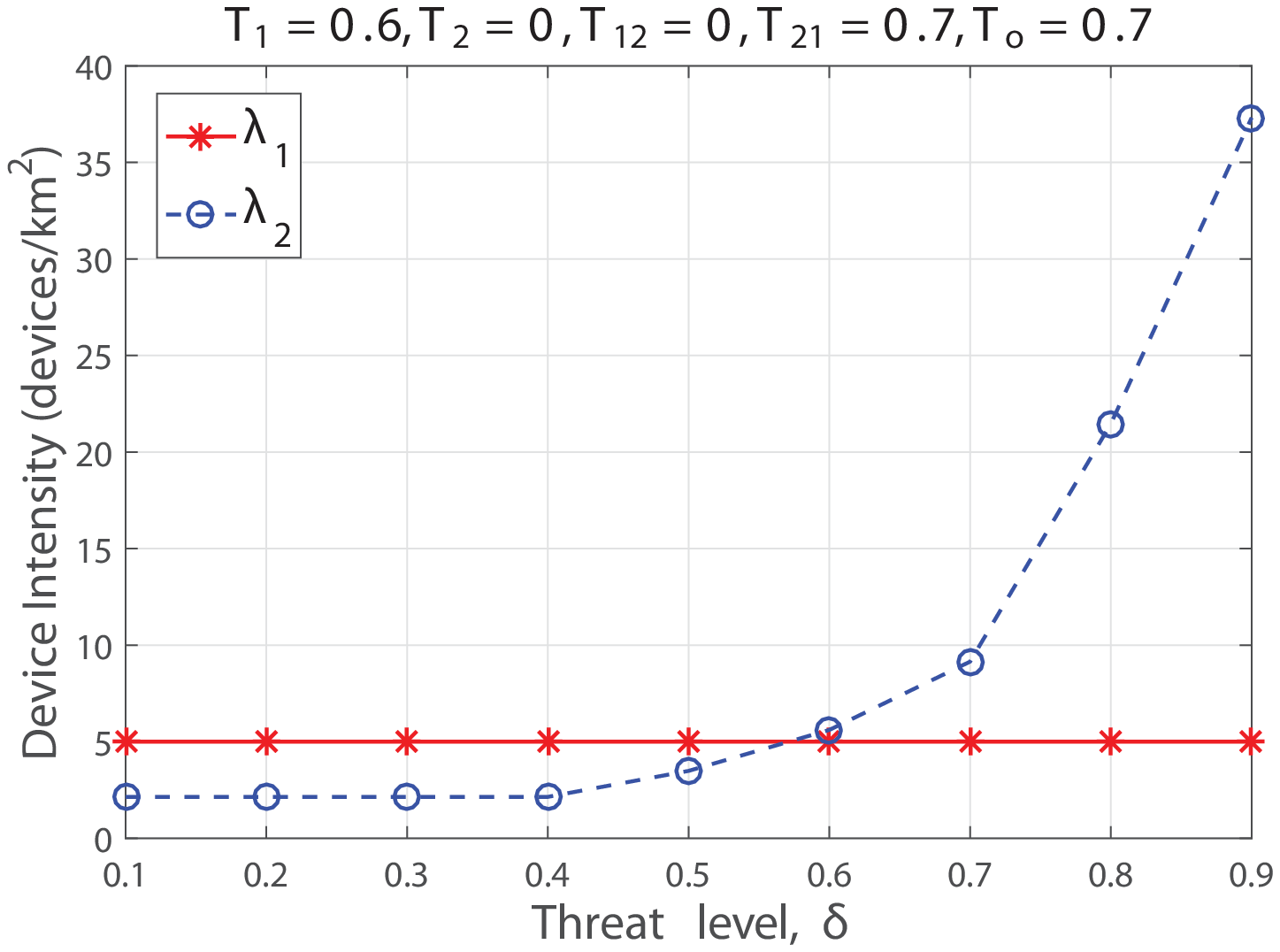}}\
\subfigure[Cost function against threat level.]{\label{enc_cost}\includegraphics[width=2.2in]{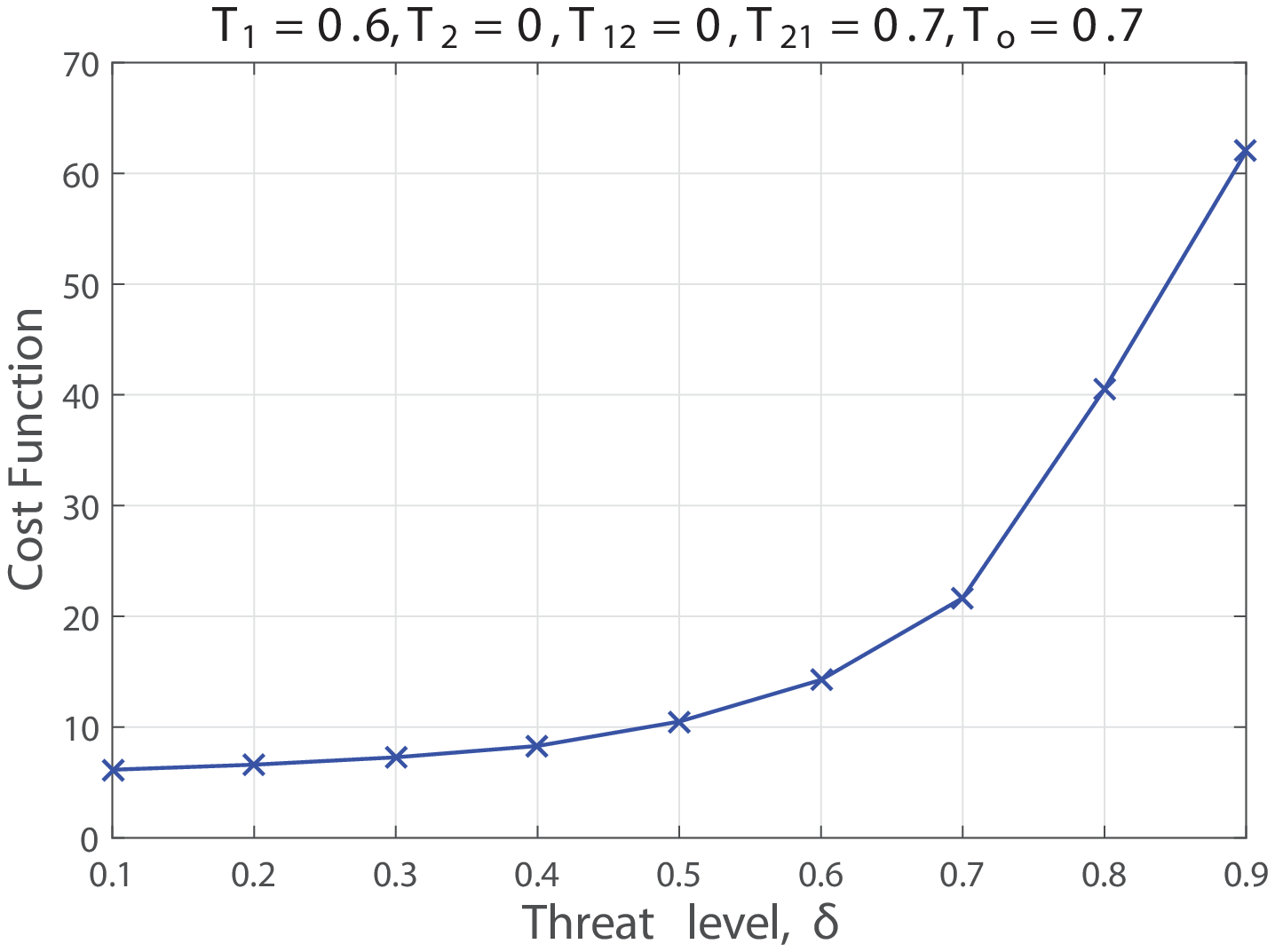}}
\end{center}\vspace{-0.0in}
\caption{Optimal network parameters for the encounter mission.}\vspace{-0.1in}
\label{Fig_encounter}
\end{figure*}

\vspace{-0.0in}
\section{Results}
In this section, we provide the results obtained by testing the developed framework under different battlefield missions. For the ease of interpretation of results, we assume a bi-layer IoBT network comprising of $M = 2$ types of battlefield things. The first type of devices is assumed to be commanders and the second type is assumed to be followers. The assumption yields a simple yet natural network configuration in a battlefield, e.g., being composed of soldiers and distributed commanding units. Let $m=1$ correspond to the commander layer of the network and $m=2$ correspond to the follower layer. The allowable physical parameter ranges of the respective devices are selected as in Table~\ref{parameters} unless otherwise stated. The parameters imply that the commanders have a higher transmission range but lower deployed density while the followers have a smaller communication range but higher deployment density. In practice, the limits can be based on tactical requirements of the missions. The weights representing the relative deployment cost are chosen to be $w_1 = 0.8$ and $w_2 = 0.2$ implying that the deployment cost of the commanding units is much higher than the follower units. The unit cost of power $p$ is selected according to the importance of each mission and the path-loss exponent $\eta = 4$.


\vspace{-0.0in}
\subsection{Mission Scenarios}
In the battlefields, there can be several types of missions such as intelligence, surveillance, encounter battle, espionage, reconnaissance, etc. In our results, we will focus particularly on the two most common mission scenarios, i.e., intelligence and encounter battle. Both of them have completely different requirements in terms of the information flow in the network, which are described as follows:

\subsubsection{Intelligence} In the intelligence mission, the goal is to provide commanders with the information from a range of sources to assist them in operational or campaign planning. It implies that there is a need for strong coordination in the follower network, i.e., soldiers and other sensors, as well as reliable flow of information from the follower network to the commander network. The coordination among commander network may not be that critical. Hence, to emulate such an intelligence mission, we select the following set of information spreading thresholds: $T_1 = 0$, $T_2 = 0.7$, $T_{12} = 0.8$, $T_{21} = 0$, and $T_o = 0.7$. The unit cost of power is selected to be high, i.e., $p = 40$, which signifies the preference of the network to spend less on power consumption during intelligence. The optimal physical parameters obtained for the intelligence mission against increasing threat level $\delta$ are shown in Fig.~\ref{Fig_intelligence}. There are several interesting observations in the intelligence mission. A general trend is that the required transmission ranges and deployment densities increases as the threat level increases. Consequently, the cost function, which signifies the deployment and operation cost of the network, also increases as shown in Fig.~\ref{intel_cost}. Fig.~\ref{intel_range} and Fig.~\ref{intel_density} show that the transmission range of the commanders is always higher than the followers while the densities of the followers is higher than that of the commanders. This observation makes sense as the followers equipped with sensors should be more in the total number while the commander network should have a larger influence area to be able to gather information for the intelligence mission. Another important observation is that the framework tends to increase the deployment density of the devices first before increasing their transmission ranges. It is due to a high cost of power consumption that tends to force the devices to minimize the transmission ranges.

\vspace{-0.0in}
\subsubsection{Encounter Battle} In the encounter battle or meeting engagement scenario, there is a contact between the battling forces. In such situations, commanders need to act quickly to gain advantage over the opponents. This requires robust information spreading from the command network to the follower network. Hence, there is a need for strong coordination among commanders and a reliable information flow from commanders to followers. Additionally, the common status information sharing between all network devices must be high to ensure accurate decision-making. Therefore, we set the following information spreading thresholds: $T_1 = 0.6$, $T_2 = 0$, $T_{12} = 0$, $T_{21} = 0.7$, and $T_o = 0.7$. In addition, the unit cost of power is selected to be low, i.e., $p = 8$, which implies a willingness of the network to spend more on power consumption during the encounter battle. Also, we fix the deployment density of the commanding devices as $\lambda_1^{\min} = \lambda_1^{\max} = 5$ km$^{-2}$ since it may not be practical to increase or decrease the number of commanders during an encounter battle. The resulting optimal parameters against the changing threat level are presented in Fig.~\ref{Fig_encounter}. In contrast with the intelligence mission, the framework tends to increase the transmission ranges of the devices first before increasing deployment, in the encounter battle. This observation is a result of the lower unit power cost for the encounter battle. Fig.~\ref{enc_range} shows that the transmission range of the commanders and followers increases as the threat level increases until the capacity of the follower devices is reached at $\delta = 0.4$. To counter higher levels of threat, the framework increases the deployment of follower nodes while the density of commanders is fixed as shown in Fig.~\ref{enc_density}. Once the threat level is higher than $\delta = 0.7$, the transmission range of the followers is actually reduced to decrease power cost as increasing the device density is directly related to the power consumption. The cost function for the encounter battle in Fig.~\ref{enc_cost} is lower than the intelligence mission in Fig.~\ref{intel_cost} mainly due to the difference in power cost.

Many other interesting mission scenarios can be emulated by defining the information thresholds as well as the physically constrained network parameters. Moreover, further insights can be obtained by investigating the behaviour of the system with more sophisticated network structure, i.e., more types of devices and their respective roles. However, in this work, we do not delve into these details since they are specific to the battlefield missions and the actual combat equipment used in the battlefields.



\vspace{-0.0in}
\section{Conclusion \& Future Work}
In this paper, we have presented a generic framework for secure and reconfigurable design of IoT empowered battlefield networks. The framework provides a tractable approach to tune the physical network parameters to achieve the desired real-time data dissemination among different types of battlefield devices according to the assigned missions. It takes into account the perceived threat level from the opponent as well as the costs involved in deployment and operation of combat equipment to provide a robust and cost effective design of communication networks in battlefields which can be highly useful in military planning. Optimized network parameters are provided for the two typical mission scenarios of intelligence and encounter battle in which the desired information spreading direction is completely opposite to one another. Results have shown that the mission goals can be achieved by either changing the deployment of combat units or by changing their transmission powers or both in response to a changing threat perception, according to the design preferences.

Although, the IoT is being widely accepted and appreciated by the commercial sector due to the huge economic impact, the military is still reluctant to adopt this technology due to the privacy and security issues. The main concern is that without proper safeguards, this linkage of systems provided by IoBT could be compromised leading to disastrous consequences. Hence, the privacy and security aspects of IoBT are one of the major challenges that need to be addressed by the researchers.

\appendices
\vspace{-0.0in}
\section{Proof of Lemma~\ref{combined_degree_lemma}} \label{combined_degree_lemma_proof}
Let $x$ denote a typical device in the combined network. The probability that device $x$ has a degree $k$ can be expressed as follows:
\begin{align} \label{joint_degree_eq}
\mathbb{P}(x\in \Phi_m)\mathbb{P}(K_o = k|x \in \Phi_m)
\end{align}
Now, $\mathbb{P}(x\in \Phi_m) = \frac{\lambda_m}{\Lambda}$ and $\mathbb{P}(K_o = k|x \in \Phi_m) = \frac{\exp(- \Lambda \pi r_m^2)(\Lambda \pi r_m^2)^k}{k!}$. Substituting these in~\eqref{joint_degree_eq} and summing over all possible values of $m$ proves the result.

\vspace{-0.0in}
\section{Proof of Uniqueness} \label{uniqueness}
To prove that the fixed point equation described by \eqref{fixed-point-equation} has a unique solution in the domain $\Theta^{(i)} > 0$, we make use of the Banach fixed-point theorem (or contraction mapping theorem)~\cite{contraction_mapping}. We prove that the functional
\begin{align}
F(\Theta(\alpha)) = \frac{1}{\mathbb{E}[K_i]}\mathbb{E}\left[\frac{K_i^2 \alpha \Theta(\alpha)}{1 + K_i \alpha \Theta(\alpha)}\right]
\end{align}
experiences a contraction for all $\Theta(\alpha) \in (0,1]$. More precisely, we prove that $|F(\Theta_1(\alpha)) - F(\Theta_2(\alpha))| \leq c|\Theta_1(\alpha) - \Theta_2(\alpha)|$ for any $\Theta_1(\alpha),\Theta_2(\alpha) \in [0,1]$, where $0 \leq c < 1$. The fact that the constant $c$ is strictly less than $1$ implies that the functional is contracted. The proof is as follows:
\begin{align}
&\left|F(\Theta_1(\alpha)) - F(\Theta_2(\alpha))\right| =\notag \\& \left|\frac{1}{\mathbb{E}[K_i]}\mathbb{E}\left[\frac{K_i^2 \alpha \Theta_1(\alpha)}{1 + K_i \alpha \Theta_1(\alpha)}\right] - \frac{1}{\mathbb{E}[K_i]}\mathbb{E}\left[\frac{K_i^2 \alpha \Theta_2(\alpha)}{1 + K_i \alpha \Theta_2(\alpha)}\right]\right|, \notag \\
&= \frac{\left|  \Theta_1(\alpha) - \Theta_2(\alpha)  \right|}{\mathbb{E}[K_i]} \mathbb{E} \left[ \frac{K_i^2 \alpha}{(1 + K_i \alpha \Theta_1(\alpha))(1 + K_i \alpha \Theta_2(\alpha))} \right].
\end{align}
To complete the proof, we need to show that
\begin{align}
\frac{1}{\mathbb{E}[K_i]} \mathbb{E} \left[ \frac{K_i^2 \alpha}{(1 + K_i \alpha \Theta_1(\alpha))(1 + K_i \alpha \Theta_2(\alpha))} \right] < 1,
\end{align}
Let $g(K_i) = \frac{K_i^2 \alpha}{(1 + K_i \alpha \Theta_1(\alpha))(1 + K_i \alpha \Theta_2(\alpha))}$. It can be proved that $g(K_i)$ is concave for $K_i \geq 0$ by showing that $g^{\prime \prime}(K_i) < 0, \forall K_i \geq 0$. Therefore, using Jensen's inequality~\cite{jensen}, we can say that $\mathbb{E}[g(K_i)] \leq g(\mathbb{E}[K_i])$, with equality iff $K_i$ is deterministic. It follows that
\begin{align}
&\frac{1}{\mathbb{E}[K_i]} \mathbb{E} \left[ \frac{K_i^2 \alpha}{(1 + K_i \alpha \Theta_1(\alpha))(1 + K_i \alpha \Theta_2(\alpha))} \right] < \notag\\
& \frac{\mathbb{E}[K_i] \alpha}{(1 + \mathbb{E}[K_i] \alpha \Theta_1(\alpha))(1 + \mathbb{E}[K_i] \alpha \Theta_2(\alpha))} = \notag \\
& \frac{1}{\Theta_1(\alpha) + \Theta_2(\alpha) + \mathbb{E}[K_i]\alpha\Theta_1(\alpha)\Theta_2(\alpha) + \frac{1}{\mathbb{E}[K_i]\alpha}    }. \label{condition_contraction}
\end{align}
The expression in~\eqref{condition_contraction} is strictly less than $1$ only if the following condition is satisfied:
\begin{align}\hspace{-0.1in}
\Theta_1(\alpha) + \Theta_2(\alpha) + \mathbb{E}[K_i]\alpha\Theta_1(\alpha)\Theta_2(\alpha) + \frac{1}{\mathbb{E}[K_i]\alpha} > 1. \label{condition_contraction2}
\end{align}
The condition in~\eqref{condition_contraction2} depends on the relative magnitudes of the quantities $\mathbb{E}[K_i]$ and $\alpha$. Regardless, it reveals that we need to exclude the values of $\Theta(\alpha)$ that are too close to zero. For sufficiently large values of $\Theta(\alpha)$, it is clear from \eqref{condition_contraction2}, that $F(\Theta(\alpha))$ is indeed a contraction with respect to the absolute value metric. Hence, by the contraction mapping theorem, $F(\Theta(\alpha))$ has a unique fixed point in the domain $\Theta(\alpha) > 0$.

\vspace{-0.15in}
\section{Proof of Existence} \label{existence}
The non-zero equilibrium solution can be obtained by solving the following equation:
\begin{align}
1 = \frac{1}{\mathbb{E}[K_i]} \mathbb{E} \left[ \frac{K_i^2 \alpha}{1 + K_i \alpha \Theta(\alpha)} \right]. \label{non_zero_eq}
\end{align}
Let $h(\Theta(\alpha)) = \frac{1}{\mathbb{E}[K_i]} \mathbb{E} \left[ \frac{K_i^2 \alpha}{1 + K_i \alpha \Theta(\alpha)} \right]$. We need to find a solution to the equation $h(\Theta(\alpha)) = 1$ in the domain $0 < \Theta(\alpha) \leq 1$. It is clear that $h(\Theta(\alpha))$ is monotonically decreasing for $\Theta(\alpha) > 0$. Therefore, it is sufficient to show that $h(0) > 1$ and $h(1) < 1$ for a unique non-zero solution to exist for the equation $h(\Theta(\alpha)) = 1$. This result is proved below:
\begin{align}
&h(0) = \frac{1}{\mathbb{E}[K_i]} \mathbb{E} \left[ K_i^2 \alpha \right] = \alpha \frac{\mathbb{E}[K_i^2]}{\mathbb{E}[K_i]}, \\
&h(1) = \frac{1}{\mathbb{E}[K_i]} \mathbb{E} \left[ \frac{K_i^2 \alpha}{1 + K_i \alpha}  \right] = \frac{1}{\mathbb{E}[K_i]} \mathbb{E} \left[ K_i\frac{K_i \alpha}{1 + K_i \alpha} \right], \notag \\ & \qquad < \frac{1}{\mathbb{E}[K_i]} \mathbb{E}[K_i] = 1.\label{h_1}
\end{align}
In~\eqref{h_1}, the inequality follows from the fact that $\frac{K_i \alpha}{1 + K_i \alpha} < 1, \forall K_i > 0, \alpha > 0$. A non-zero solution to \eqref{non_zero_eq} exists only if $h(0) \geq 1$, which implies that $\alpha \geq \frac{\mathbb{E}[K_i]}{\mathbb{E}[K_i^2]} $.

\vspace{-0.1in}
\section{Proof of Theorem~\ref{theorem_solution}} \label{solution}
\begin{figure}
  \centering
  \includegraphics[width=2.8in]{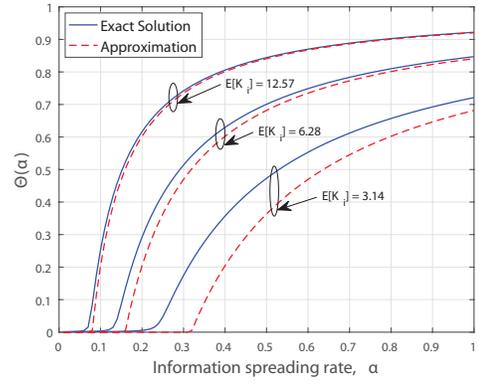}\\
  \caption{Accuracy of Jensen's lower bound. \vspace{-0.4cm}}\label{approx_accuracy}
\end{figure}

Obtaining the non-zero solution for the fixed point equation \eqref{fixed-point-equation} in closed form is not possible since we need to solve the following equation for $\Theta(\alpha)$:
\begin{align}
1 = \frac{1}{\mathbb{E}[k]} \sum_{k=0}^{\infty} \frac{K_i^2 \alpha}{1 + K_i \alpha \Theta(\alpha)} P(K_i), \label{fixed_point_exact}
\end{align}
where $P(K_i) = \frac{e^{-\lambda_i \pi r_i^2}(\lambda_i \pi r_i^2)^K_i}{K_i !}$.
Therefore, we resort to find an approximation for the solution which is asymptotically accurate. Let $g(K_i) = \frac{K_i^2 \alpha}{1 + K_i \alpha \Theta(\alpha)}$. Since $g^{\prime \prime}(K_i) > 0, \forall K_i \geq 0$, so $g(K_i)$ is a convex function for $K_i \geq 0$. Using Jensen's inequality, we can say that $\mathbb{E}[g(K_i)] \geq g(\mathbb{E}[K_i])$, with equality only if $K_i$ is deterministic. This implies the following:
\begin{align}
\mathbb{E} \left[ \frac{K_i^2 \alpha}{1 + K_i \alpha \Theta(\alpha)} \right] > \frac{\mathbb{E}[K_i]^2 \alpha}{1 + \mathbb{E}[K_i] \alpha \Theta(\alpha)}.
\end{align}
Therefore, we can write~\eqref{fixed_point_exact} as follows:
\begin{equation}
1 > \frac{\mathbb{E}[K_i] \alpha}{1 + \mathbb{E}[K_i] \alpha \Theta(\alpha)},
\end{equation}
which leads to the final solution,
\begin{align}
\Theta(\alpha) > 1 - \frac{1}{\alpha \mathbb{E}[K_i]}.
\end{align}
Using our prior knowledge that $\Theta(\alpha) \geq 0$, we need to ensure that $\alpha \mathbb{E}[K_i] \geq 1$. In general, the complete solution can be expressed as $\Theta(\alpha) > \max(0, 1 - \frac{1}{\alpha \mathbb{E}[K_i]})$.
To measure the accuracy of this bound, we solve the fixed-point equation exactly using an fixed-point iteration and compare the results for different values of $\alpha$ and $\mathbb{E}[K_i] = \lambda_i \pi r_i^2$. We choose a fixed $r_i = 0.2$ km and $\lambda_i = [25, 50, 100]$ km$^-2$, which results in $\mathbb{E}[K_i] = [3.14, 6.28, 12.57]$. A plot of the results is provided in Fig.~\ref{approx_accuracy}. It can be observed that the lower bound obtained using Jensen's inequality is tight for all values of $\alpha$ when $\mathbb{E}[K_i] \gg 1$. 

%

\vspace{-0.2in}
\bibliographystyle{IEEEtran}
\bibliography{references}

\end{document}